Correlation between bulk thermodynamic measurements and the low temperature resistance plateau in $SmB_6$

W. A. Phelan,[1,2] S. M. Koohpayeh,[2] P. Cottingham,[1,2] J. W. Freeland,[3] J. C. Leiner,[4] C. L. Broholm,[2] and T. M. McQueen[1,2*]

[1]*Department of Chemistry, Johns Hopkins University, Baltimore, Maryland 21218, USA*
[2]*Institute for Quantum Matter, Department of Physics and Astronomy, Johns Hopkins University, Baltimore, Maryland 21218, USA*
[3]*Advanced Photon Source, Argonne National Laboratory, Argonne, Illinois 60439, USA*
[4]*Quantum Condensed Matter Division, Oak Ridge National Laboratory, Oak Ridge, Tennessee 37831, USA*
[*]*mcqueen@jhu.edu*

**Abstract**

Topological insulators are materials characterized by dissipationless, spin-polarized surface states resulting from non-trivial band topologies. Recent theoretical models and experiments suggest that $SmB_6$ is the first topological Kondo insulator, in which the topologically non-trivial band structure results from electron-electron interactions *via* Kondo hybridization. Here, we report that the surface conductivity of $SmB_6$ increases systematically with bulk carbon content. Further, addition of carbon is linked to an increase in n-type carriers, larger low temperature electronic contributions to the specific heat with a characteristic temperature scale of T* = 17 K, and a broadening of the crossover to the insulating state. Additionally, X-ray absorption spectroscopy shows a change in Sm valence at the surface. Our results highlight the importance of phonon dynamics in producing a Kondo insulating state and demonstrate a correlation between the bulk thermodynamic state and low temperature resistance of $SmB_6$.

## I. Introduction

Topological insulators (TIs) are materials in which a non-trivial topology of the bulk band structure gives rise to spin-momentum locked surface states at interfaces with normal insulators [1,2]. In addition to being spin-polarized, the surface states are protected from scattering by non-magnetic disorder. These remarkable characteristics make TIs ideal building blocks for breakthroughs in many fields, including spintronics and quantum computation [3]. The theory and experimental consequences of a non-trivial band topology in weakly correlated, inversion symmetric materials such as $Sb_2Te_3$, $Bi_2Te_3$, and $Bi_2Se_3$ are now well established [4-6].

While topologically protected surface states in strongly correlated materials are an exciting possibility with theoretical backing, the experimental situation is less clear. $SmB_6$ is a long-studied Kondo insulator [7,8] which has come back into focus following the theoretical predictions that it may harbor topologically protected surface states below its Kondo temperature ($T_K$) due to the interplay of spin orbit coupling with strong electron-electron interactions [9-11]. Most samples of $SmB_6$ show a resistance plateau at low temperature (*ca.* 5-10 K). Previous literature hypothesized that this low temperature remnant metallicity originates from in-gap impurity states due to imperfect sample stoichiometry and secondary phases [12,13]. More recently, well-designed electrical transport measurements performed using flux grown single crystals of $SmB_6$ have shown that the low temperature resistivity plateau is due to surface, not bulk, conduction, and may be topological in origin [14-17]. An extensive range of additional measurements, including normal and spin-resolved angle resolved photoemission spectroscopy (ARPES) [18-24], surface circular dichroism [20], scanning tunneling microscopy (STM) [25,26], and torque magnetometry [27], provide additional evidence for surface states.

Here we report the results of both surface and bulk physical properties measurements on single crystals of $SmB_6$ prepared *via* the floating zone (*FZ*) technique, both undoped and containing small levels of aluminum and carbon. We find that greater residual conduction at low temperature (i.e. more plateaued resistance) correlates with increased carbon content. Further, we find that the level of carbon systematically affects the Seebeck coefficient and specific heat. In addition to resolving a number of discrepancies in previous $SmB_6$ studies, these results establish a correlation between the bulk thermodynamic state and the surface conduction of $SmB_6$. Finally, we provide evidence through X-ray absorption spectroscopy (XAS) and X-ray magnetic circular dichroism (XMCD) that the surface of *FZ* grown $SmB_6$ is paramagnetic and that the valence state of Sm differs between the surface and the bulk.

**II. Methods**

Single crystals of $SmB_6$ with approximate dimensions of 30 mm in length and 6 mm diameter were grown from rods of polycrystalline $SmB_6$ (Testbourne Ltd, 99.9%) using a four-mirror optical floating zone (*FZ*) furnace (Crystal Systems Inc FZ-T-12000-X-VPO-PC) with 4 × 3 kW Xe lamps. For undoped $SmB_6$, separate crystals were prepared with $SmB_6$ from Testbourne Ltd (99.5%) and Alfa Aesar (AA 99.9%) and found to exhibit similar low temperature resistances. In order to simulate an "aluminum flux" growth of $SmB_6$, a pressed pellet with a targeted mass of *ca*. 0.20 g was prepared with a $SmB_6$ (AA 99.9%):Al (AA 99.97%) mixture in a ratio of 50:50 wt %. The prefabricated $SmB_6$/Al pellet was placed between the seed and the feed rods and passed through the molten zone during the growth procedure. For the carbon incorporated growth, the same procedure was followed with a pressed pellet prepared from a 0.5 g pre-homogenized $SmB_6$ (AA 99.9%):C(AA 99.9995%) mixture in a ratio of 15:85 wt %. This pellet was arc-melted, turned over, and remelted in a Zr-gettered argon atmosphere to ensure homogeneity. The mass loss of the resultant boule was determined to be negligible. For all growths, the molten zone was passed upwards with the feed rod being at the top and the seed rod below it and the pellets. One pass was sufficient to incorporate aluminum and carbon throughout each grown crystal. The mirrors were held in fixed positions and both rods were translated downwards. Growths were carried out under flowing ultra-high purity Ar at a pressure of 2 bar and a flow rate of 2 L/min. The rotation rates for the seed (growing crystal) and the feed rod were 10 and 3 rpm, respectively. The overall growth rate for both experiments was 10 mm/h. Only one zone pass was performed for each growth. Slices of the crystals, close to the [100] orientation, were cut using a diamond saw.

Laboratory powder X-ray diffraction was performed using Cu $K\alpha$ radiation on a Bruker D8 Focus diffractometer with a LynxEye detector. A Si standard, with lattice parameter $a$ = 5.43102 Å, was used in all samples to obtain accurate relative unit cell parameters. Rietveld refinements were performed in TOPAS (Bruker AXS) to extract lattice parameters of $a$ = 4.1342(1) Å and $a$ = 4.1338(1) Å near the beginning and end of the growth for $SmB_6$:C, respectively.

To prepare a single crystal of $LaB_6$, La (AA 99.9%) and B (AA 99%) were combined and pressed into 0.73 g pellets with a 15% mass excess of B. These pellets were individually arc-melted using the same procedure described for $SmB_6$/C pellet and then combined and ground using a stainless steel mortar and pestle. The resulting powder was pressed into a rod using a hydrostatic press. This rod was divided into two pieces, and the resulting rods were heated in a high temperature vacuum furnace at 5º C/hr to 1600 ºC and held for 8 hours. The sintered rods were used to grow a single crystal of $LaB_6$ using the same experiential growth parameters described above for the $SmB_6$ growths.

An Amray 1810 scanning electron microscope (SEM) was used to collect images of the $SmB_6$/C samples. An accelerating voltage of 30 kV and a beam current of 40 $\mu$A were used to collect images of naturally cleaved sample surfaces. Images were collected using both secondary and backscattering modes.

Resistance and Seebeck data were obtained using the resistivity and thermal transport options of a 9-Tesla Quantum Design Physical Property Measurement System (PPMS). Unless otherwise stated, all measurements were performed using standard four-probe techniques, with approximate sample dimensions of length ~ 1 mm and cross sectional area ~ 3 mm (thickness ~ 0.8 mm). Excitation currents for the resistance measurements were set sufficiently low to avoid significant Joule heating (see Fig. S5). Pt leads were used for the resistance measurements, and Au plated Cu leads were used for Seebeck measurements. The leads were mounted onto the samples using silver epoxy. All reported specific heat data were collected using the semi-adiabatic pulse technique as implemented for the PPMS.

The XAS/XMCD experiments were conducted at beam line 4-ID-C of the Advanced Photon Source located at Argonne National Laboratory. Surface-sensitive spectra were collected using total electron yield and bulk sensitive data with partial fluorescence yield using circularly polarized X-rays with a grazing 10º configuration. The XMCD spectra are given by the difference between the absorption spectra of the right and left circularly polarized X-rays. Measurements were performed at T = 10 K with an applied magnetic field of $\mu_0 H = 5\ T$ in the plane of the sample.

**III. Results and Discussion**

Fig. 1(a) shows the temperature dependent resistance data from a high resistance pure $SmB_6$ sample, with a thickness of 1000 μm. This sample has a resistance that continues to increase down to at least T = 2.7 K (R > 40 Ω at T = 2.7 K, $R/R_{300K}$ > 50,000, R□ > 100 Ω). No plateau is observed below *ca.* 5 K for this sample, unlike flux grown samples of $SmB_6$ [8,12,14-16]. To test for a thickness dependence of the resistance behavior, as has previously been observed, Fig. 1(b) shows the resistance data of $SmB_6$ from a different growth, measured with two different thicknesses, 800 µm (non-thinned) and 48 µm (thinned). Again, the resistance continues to diverge to the lowest temperatures measured (T = 2 K), although there is a clear "knee" at *ca*. 5K. If the conduction for the non-thinned and thinned samples is three dimensional, then $\rho_{non-thinnned} = \rho_{thinned}$ or $R_{non-thinned}\frac{(A_{non-thinned})}{L_{non-thinned}} = R_{thinned}\frac{(A_{thinned})}{L_{thinned}}$. Therefore

$\frac{R_{non\text{-}thinned}}{R_{thinned}}$ should equal a constant factor. Fig. 1(c) shows this ratio of the resistances. It deviates from a constant value at a temperature close to the "knee". This is consistent with previous reports which have interpreted the change in slope as a switch from bulk to surface dominated conduction [14,15]. Unlike previous reports, however, we do not observe true plateauing of the low temperature resistance. Instead, a two channel fit suggests a small semiconducting gap of 16(4) K for the low temperature channel (Table SI).

To test if aluminum, the flux typically used to grow $SmB_6$, is the origin of a more significant plateau, a set of samples from a simulated "aluminum flux" growth were prepared. Normalized resistance data for these samples compared to pure $SmB_6$ are shown in Fig. 2(a). Samples grown in the presence of aluminum show a substantial reduction in the normalized resistance. Further, there is a gradual plateauing below T = 5 K. However, there is no systematic trend with respect to position of the slices cut from the $SmB_6$/Al crystal. These observations are consistent with filamentary inclusions of aluminum that provide a low resistance pathway and "short out" the insulating bulk at low temperature.

However, incorporation of aluminum does not produce the dramatic resistance plateaus well-known in this system. The difference between our *FZ* grown crystals and those previously reported provides a possible explanation. Previous $SmB_6$ single crystals prepared *via* the *FZ* procedure have been reported to exhibit a low temperature resistance plateau [13,28,29]. The most recent work that describes the crystal growth in detail, states that the carbon based binders PVA and PVB were used to prepare the rods necessary for the *FZ* growth [30], whereas no binder was used during our preparations. Further, carbon is also a common impurity in both boron and aluminum, and is difficult to quantify analytically because most methods have carbon present elsewhere (e.g., to affix the sample to the stage for energy or wavelength dispersive X-ray spectroscopy).

To test whether the presence of carbon might be influential in producing a low temperature resistance plateau, we prepared $SmB_6$ crystals containing carbon. The temperature dependent normalized resistance data for these samples are shown in Fig. 2(b). The most remarkable observation is the systematic appearance of a resistance plateau below T = 6 K, which becomes less prominent as a function of passed molten zone position. This observation suggests that the carbon content was varied systematically throughout the SmB6/C crystal. To test whether this plateau has the same origin in surface conductivity as previously reported [31], a measurement was performed on sample **5** before it was thinned (**5'**). Another resistance measurement was performed using sample **5** by polishing **5'** without removing the electrical contacts, Fig. 2(c). The resistance values at 2 K and 300 K for the two samples change by 1.9% and 26%, respectively. This demonstrates that the low temperature plateau is independent of sample thickness, which is consistent with surface-dominated conduction at low temperature.

These results show that this well-known yet not well-understood feature can be controlled *via* chemical modification, specifically by adding carbon.

Suspecting that microstructural differences might be responsible for the dependence of surface conduction in carbon-containing $SmB_6$, scanning electron microscope (SEM) micrographs (not shown) on naturally exposed surfaces of **1-5** were collected. No observable changes in the microstructure between samples were observed down to *ca*. 50 nm.

To further elucidate the origin of the dependence of the normalized resistance plateau on carbon content, we performed Seebeck, magnetization, and specific heat measurements. The results of our Seebeck measurements are similar to those previously reported for $SmB_6$, both *FZ* and flux grown, where the sign of the Seebeck coefficient is negative (positive) below (above) T = 300 K [32,33]. The temperature at which the Seebeck coefficient crosses zero corresponds to the point where thermally generated carriers cancel the effect of chemically doped carriers. As is shown in the inset of Fig. 3, there is a small but significant (Fig. S1) shift of this cancellation to higher temperatures for different samples along the length of the $SmB_6$/C crystal. This supports the conclusion that carbon results in the addition of bulk n-type carriers, expected if carbon replaces boron:

$$B_B^x + C \longrightarrow C_B^{\bullet} + e^- + B \qquad (1)$$

This also suggests that the added carbon content is proportional to the magnitude of the numbers (i.e. the larger the number, the greater the carbon content.).

Magnetization measurements also support a systematic change in electron count on addition of carbon. The magnetization data as a function of temperature is shown in Fig. 4(a). A Curie-Weiss analysis of the high temperature (100 – 300 K) region yields an effective Curie constant that increases with sample number, Fig. 4(b) (Table SII). This, along with the Seebeck data, suggests the overall electron count is being altered upon carbon substitution.

Fig. 5(a) shows the specific heat collected on pure and carbon-containing $SmB_6$ specimens (**2** and **5**), as compared with isostructural, non-magnetic $LaB_6$. In all cases, a large excess heat capacity is observed. This corresponds to a large additional change in entropy in $SmB_6$ that is not present in $LaB_6$. The point of maximum entropy change occurs at T ~ 40 K, close to the temperature below which the resistance sharply increases. This change in entropy can be associated with the formation of the Kondo insulating state in $SmB_6$. The corresponding Kondo temperature is then estimated to be $T_K$ = 80 – 400 K, depending on the approximation used to relate the peak in $C_p T^{-1}$ to $T_K$, in good agreement with previous literature [18-24]. By integrating the difference in $C_p T^{-1}$ between $SmB_6$ and $LaB_6$, it is possible to estimate the total entropy associated with this crossover (Fig. 5(b)). We find that the entropy change is $\Delta S \sim 23$ J $K^{-1}$ mol-f.u.$^{-1}$ = 2.9R (R = 8.314 J $K^{-1}$ mol-f.u.$^{-1}$), with the most stringent lower limit being $\Delta S > 19$ J $K^{-1}$ mol-f.u.$^{-1}$ = 2.3R depending the details of how the phonon background is subtracted using the non-magnetic analog $LaB_6$. This is substantially larger than the $\Delta S_{CF} \sim 0.4R$ expected based on the observed crystal field excitations in neutron scattering on powder samples that suggest only one thermally accessible set of crystal field levels at ~28 meV [34]. It is also greater than $\Delta S_{MV} \sim 2.0R$ expected for a $Sm^{2+}/Sm^{3+}$ mixed valence system in which the ratio is fixed at 1:1, as X-ray absorption spectroscopy (XAS) and Sm-Mössbauer studies imply [35,36], and the accessible J multiplets are J = 0, 2, and 3 for $Sm^{2+}$ and J = 5/2 for $Sm^{3+}$, as previously suggested [7]. Some of the excess entropy will be due to the localization of the conduction electrons when the hybridization gap forms, but a more likely explanation for the large value of $\Delta S$ we observe is an additional phonon (lattice) contribution. This is similar to what has been observed by neutron scattering in $YbB_{12}$ [37]. Further, a soft phonon mode is known to exist in $SmB_6$ [38], and unusual changes in atomic displacement parameters and lattice parameters are known crystallographically [39,40]. This implies that understanding the phonon dynamics is essential to understanding $SmB_6$. It also implies that understanding phonon dynamics will be critical in elucidating any topological properties of related hexaborides, e.g. $YbB_6$, as has recently been suggested [41-43].

There are two noticeable changes in the specific heat data as a function of the $SmB_6$/C sample. First, at temperatures in the range of the gap opening (T ~ 40 K) the peak broadens. This is consistent with a systematic addition carbon of defects found along the grown crystal, and these defects have the effect of disrupting formation of the Kondo lattice. Second, there is an upturn in $C_p T^{-1}$ at T < 5 K that becomes more prominent, shown in more detail in Fig. 6(a). In each case, this upturn is sensitive to an applied magnetic field. We found that all specific heat data sets could be fit well to the expression:

$$C_p = \gamma T + \beta_3 T^3 + AT^3 \ln(T/T^*) + BT^{-2} \quad (2)$$

The first and second terms are standard electronic and lattice contributions to the specific heat, respectively. The third term is typically associated with an exchange enhanced paramagnetic metal [44], but is a correction to the T-linear electronic term that can occur in any Fermi liquid with momentum dependent quasiparticle interactions [45], or electron-phonon coupling [46]. Such a term has been theoretically [47] and experimentally found to occur in heavy fermion systems [44]. Further, the heat capacity of $SmB_6$ measured in a previous report shows this same $AT^3\ln(T/T^*)$ dependence down to T ~ 0.1 K [13]. The fourth term, $BT^{-2}$, is the high temperature expansion of a Schottky anomaly, and is only a minor contribution to specific heat, below T ~ 2.8 K (Fig. S3). To aid in the data fitting procedure, Eq. 2 was recast in the following form:

$$C_p T^{-1} = \gamma + \beta T^2 + AT^2 \ln(T) + BT^{-3} \qquad (3)$$

with $\beta = \beta_3 - A\ln(T^*)$. The fits over the temperature range T = 2 to 10 K are drawn in Fig. 6(a) and the resulting coefficients tabulated in Table SIII.

Remarkably, the β and A parameters from all samples (with and without carbon) lie on a universal curve, Fig. 6(b), with a single $\beta_3$ lattice contribution and a single characteristic temperature of -ln(T*(K)) = -2.8 (T* = 17 K) associated with Kondo hybridization. Test fits of the pure and $SmB_6$/C-5 samples over different temperature ranges (T = 3 to 12 K, 2 to 12 K) showed that while there is some variability in the precise values of ß and A depending on the range used, the values still lie on this universal curve. The $\beta_3$ lattice contribution extracted in this way corresponds to a Debye temperature of $\theta_D$ = 230 K (we have followed the assumption typically made in the hexaboride literature, which assumes one atom per mole of material (i.e. per mol (La/Sm)$B_6$) [48]). This is in good agreement with the previously reported $\theta_D$ ~ 250 K of $LaB_6$ which corresponds to a local La mode [48].

Fig. 7(a) shows key parameters from the Seebeck and specific heat measurements plotted versus $1/E_a$, a measure of the degree of plateauing of the low temperature resistivity, defined as $1/E_a = \lim_{T\to 0}\left[1/T \ln\left(R/R_{300K}\right)\right]$ (Fig. S4). The negative to positive Seebeck crossover, $T_{s=0}$ (left), the coefficient γ, and temperature of the minimum value of the $C_p/T$ (right) increase concomitantly with $1/E_a$. This demonstrates a correlation between the bulk electronic state and surface conductivity in $SmB_6$ (surface contributions to the specific heat constitute ≪ 1% of the observed values).

Fig. 7(b) diagrams a possible mechanism by which changes in the bulk can have such a significant impact on surface conduction. The density of states (DOS) in a Kondo insulator for $T \ll T_K$ consists of two sharp peaks, from hybridization of localized rare earth f states and conduction states, separated by a small gap. Our specific heat data can be used to estimate the Fermi level ($E_F$) of each sample on such a plot in the following way. First, the magnitude of the $\gamma T$ term is proportional to the DOS at $E_F$, so samples with larger $\gamma$ are located in regions of higher DOS. Second, the magnitude of the thickness-independence low-temperature resistance is higher when the band filling of surface states is lower. This places the Fermi level of undoped $SmB_6$ in or close to the gap, while the carbon containing samples have $E_F$ beyond the point along the band edge where the curvature changes from positive to negative. The $E_F$ positions estimated in this way are drawn as vertical dashed lines, and are consistent with the observed increase in n-type doping as carbon is added. If surface states are present in the neighborhood of the gap, the position of $E_F$ in the bulk acts to pin the band filling of the surface states, in turn modulating their conductivity. Critically, for undoped *FZ* $SmB_6$, the bulk Fermi level is positioned such that the carrier density at the surface is low.

What, then, is the origin of this low temperature resistance plateau in $SmB_6$? One possibility is that the plateau is a consequence of Joule heating of the sample. This explanation is unlikely, however, as resistance values obtained from zero-current extrapolation of IV curves agree with our low-excitation current measurements (see Fig. S5, S6, and S7). Instead, in agreement with previous interpretations, a far more likely explanation is that it is due to surface conduction. Theoretical arguments suggest that surface states exist due to a non-trivial bulk band topology [9-11]. A second potential source of surface conduction, namely “normal” electronic surface states associated with the boron framework. A third possibility is a chemical change at the surface: it is well-known that the surfaces of metal hexaborides can become metal deficient, especially in the case of $SmB_6$ where surface vacancy patterns have been observed in low energy electron diffraction and helium ion scattering spectroscopy [49]. This loss of Sm (and corresponding oxidation of some surface $Sm^{2+}$ to $Sm^{3+}$) naturally disrupts the Kondo lattice and coherence [50]; a fourth possibility is that surface reconstructions, such as those observed by STM [25,26], change the electronic structure at the surface. A fifth possibility suggested by our data, especially the ability to have the resistance plateau occur over a wide range of absolute resistance values and the previously reported dependence of the low temperature plateau on cleaning and etching procedures, is that both topological and non-topological surface states are simultaneously present. The precise crystallographic orientations of surfaces where the normal and non-trivial surface states may coexist depends on the individual origins of each state. For example, if the normal surface states are associated with the boron framework and $SmB_6$ belongs to the class of strong $Z_2$ topological insulators, then both normal and non-trivial states will coexists on all surfaces. However, if normal surface states are associated with the boron framework and $SmB_6$ is a topological crystalline insulator, as has recently been predicted [51], then the coexistence of these states is limited to the (110) surface. Finally, if it is only polarity driven surface reconstructions that give rise to normal surface states and $SmB_6$ belongs to the class of strong $Z_2$ topological insulators, then the surface of coexistence is the (100). A sixth possibility is that $SmB_6$ is poised at the boundary between normal and topological states and small changes in electron count push the system to one side or the other of the phase boundary. The estimated Sm valence of 2.5+, from XAS and Mössbauer measurements [35,36], is very close to the theoretical boundary of 2.56+ [52], and this scenario would reconcile our results with theory and with previous experiments.

To understand the nature of the $SmB_6$ surface in more detail, we collected Sm-$M_{4,5}$ XAS and XMCD spectra simultaneously in surface (electron yield) and bulk (fluorescence yield) modes at T = 10 K and $\mu_0H = 5$ *T* (Fig. 8(a) and 8(b)). The bulk spectra are consistent with a mixture of $Sm^{2+}$ and $Sm^{3+}$, with no appreciable magnetization, as previously reported [35,53]. In contrast, the surface consists of almost entirely $Sm^{3+}$ and shows a discernible XMCD signal characteristic of a net magnetic moment. The magnitude of the surface XMCD response is small, approximately 1/10$^{th}$ that observed in ferromagnetic $Sm_{0.974}Gd_{0.02}Al_2$ [54]. Using sum rules [55,56], the estimated orbital ($M_L$) and spin ($M_S$) moments of $Sm^{3+}$ at the surface are 0.09 $\mu_B$ and -0.05 $\mu_B$, respectively. These values correspond to a net magnetic moment of 0.09 $\mu_B$ at T = 10 K and $\mu_0H$ = 5 *T*, in good agreement with the 0.08 $\mu_B$ expected for paramagnetic $Sm^{3+}$ ions, implying the surface is paramagnetic at T = 10 K and distinct from the bulk Kondo insulator for which no induced magnetic moment under these thermomagnetic conditions by XMCD.

The surface versus bulk comparison of the XAS data is definitive proof of changes in the electronic structure at the surface of $SmB_6$. The disparate valences at the surface and bulk could give rise to a surface-to-bulk electric field and hence band bending at the interface, and would - provide one possible explanation for the dependence of surface conduction on the bulk properties, but experiments such a liquid gating are needed to clarify this issue. Further, the non-zero surface XMCD indicates magnetic ions are present on the surface. A particularly intriguing possibility that would explain the results of recent magnetotransport measurements [57] and the low frequency conduction observed by terahertz spectroscopy [58] is that the surface of undoped $SmB_6$ becomes magnetically ordered at even lower temperatures. Even if topologically protected surface states are present, such magnetic order would open a small gap in the surface density of states and remove the low temperature surface conductivity. The action of carbon would then be to suppress the formation of magnetic order (e.g., by conversion of $J \neq 0$ $Sm^{3+}$ to $J = 0$ $Sm^{2+}$ at the surface by pinning to the bulk Fermi level) and restore time reversal symmetry.

This scenario is consistent with two channel fits of the resistance data, where the primary action of carbon is to reduce the effective activation gap of surface conduction, with no significant change in the magnitude of the bulk activation gap (Table SI).

**IV. Conclusions**

We find that there is a correlation between quantities obtained from bulk thermodynamic measurements and surface conduction in $SmB_6$. Small aluminum inclusions do not dramatically affect the low temperature surface conduction in *FZ* grown samples. The $\Delta S \sim 23$ J $K^{-1}$ mol-f.u.$^{-1}$ entropy change through the Kondo coherence crossover at T ~ 40 K is too large to originate solely from magnetic and electronic degrees of freedom, and implies consideration of lattice vibrations and their coupling to magnetic and electronic states is critical to understanding $SmB_6$. Further, the valence of Sm at the surface is found to be distinct from the bulk, indicative of a change in electronic structure at the surface. In short, our results demonstrate a correlation between bulk thermodynamic properties of $SmB_6$ and the low temperature resistivity plateau arising from surface-dominated conduction, a necessary pre-requisite for topological Kondo insulating behavior.

**Acknowledgements**

We thank J. Checkelsky, N.P. Armitage, and O. Tchernyshyov for useful discussions; C.L. Chien for providing some sample materials; N. Laurita for sample manipulations; and N. Hartman for assistance with the SEM data collection. The work at IQM was supported by the U.S. Department of Energy, office of Basic Energy Sciences, Division of Material Sciences and Engineering under Grant No. DE-FG02-08ER46544. TMM acknowledges support of the David and Lucile Packard Foundation. Work at Argonne National Laboratory and use of the Advanced

Photon Source was supported by the U.S. Department of Energy, Office of Basic Energy Sciences under Contract No. DE-AC02-06CH11357.

FIG. 1. (a) R(T) data for the highest resistance $SmB_6$ samples, thickness 1.01 mm. It was not possible to measure below T = 2.7 K due to Joule heating effects. The line in the inset is a guide to the eye. (b) R(T) data for typical $SmB_6$ specimens, with thickness of ~800 μm ($SmB_6^{'}$) and ~48 μm ($SmB_6$). The inset shows that the resistances continues to diverge to the lowest temperatures measured (2 K), although there is a clear “knee” at *ca*. 5 K (c) Resistance of a non-thinned sample of $SmB_6$ ($R_{SmB6'}$) normalized by the resistance of a thinned sample of $SmB_6$ ($R_{SmB6}$) versus temperature. There is a deviation from a constant value at low temperatures, suggesting a switch away from three dimensional dominated conduction below *ca.* 5 K.

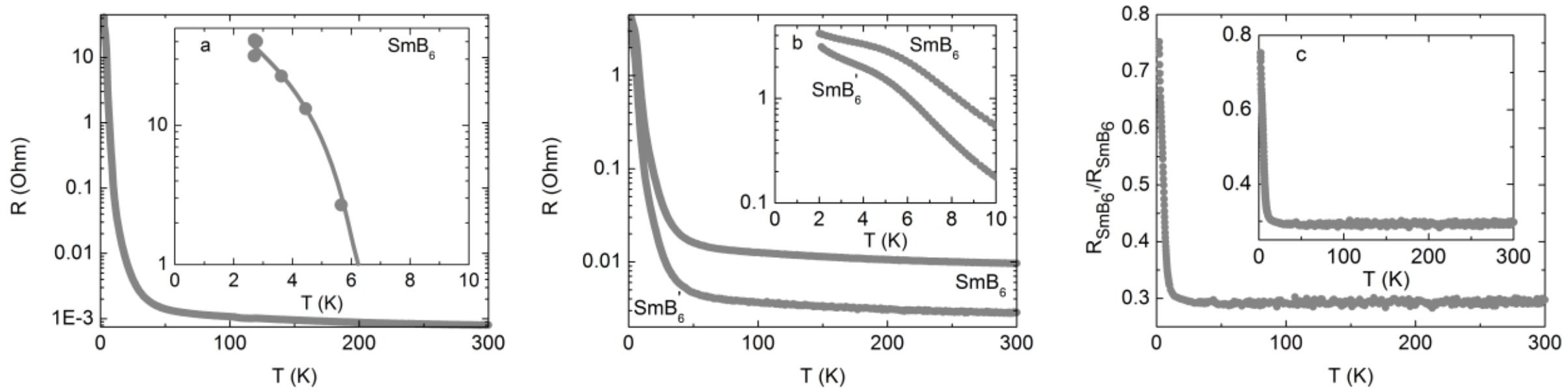

FIG. 2. Normalized resistance measurements for (a) $SmB_6$/Al and (b) $SmB_6$/C. Both insets highlight the $R/R_{300K}$ data below T = 10 K. The colored numbers in the insets are arbitrary designations representing the position of the crystal slice along the $SmB_6$/Al and $SmB_6$/C crystals (MZ = Molten Zone). All samples were of comparable geometry (see Methods, ±10% each dimension) to minimize sample shape effects that might arise due to a mixture of surface and bulk conductivity. The slope of the low temperature resistance changes systematically with the position of the slice cut along the $SmB_6$/C crystal, but not with aluminum position, indicating that variation of the bulk carbon content can be used to systematically tune the low temperature resistance plateau. The gray line shows the normalized resistance of undoped *FZ* $SmB_6$ in (a) and is off the scale of the inset in (b). (c) A comparison of the high and low temperature resistance of thinned (5) and a non-thinned (5') sample $SmB_6$/C-5. The resistance values at 2 K and 300 K for the two samples change by 1.9% and 26%, respectively. This demonstrates that the low temperature plateau is independent of sample thickness, which is consistent with surface conduction. The sample was thinned without removal of the measurement leads, so all factors except thickness were held constant.

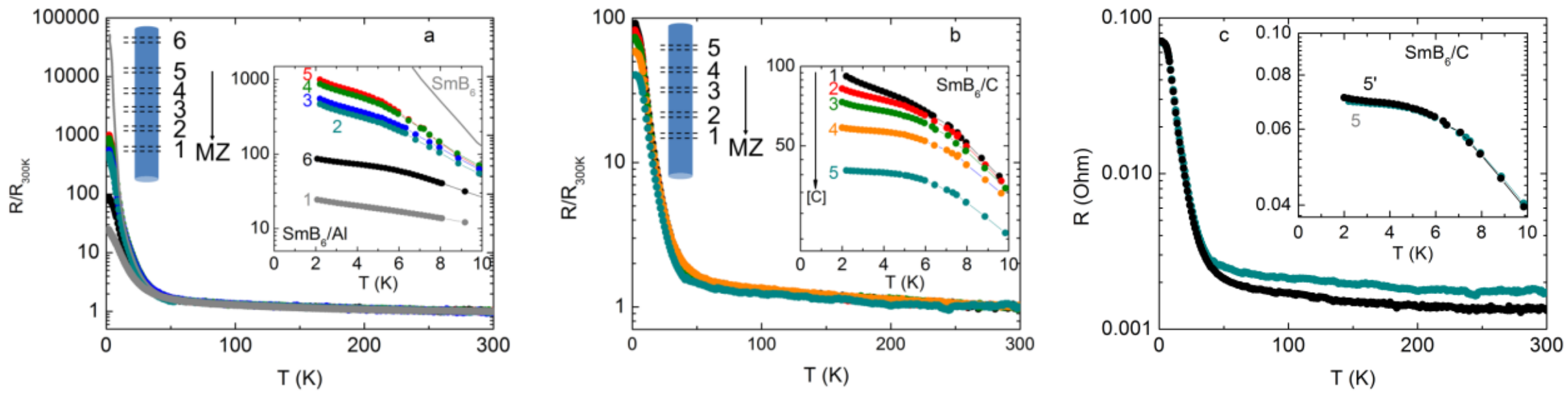

FIG. 3. Seebeck (S) coefficient vs. temperature for carbon-containing samples **1**, **3**, **4**, and **5**. $T_{S=0}$, the temperature at which thermally excited carriers produce a change in the sign of S from negative to positive upon heating, increases with position of $SmB_6$/C slice along the crystal.

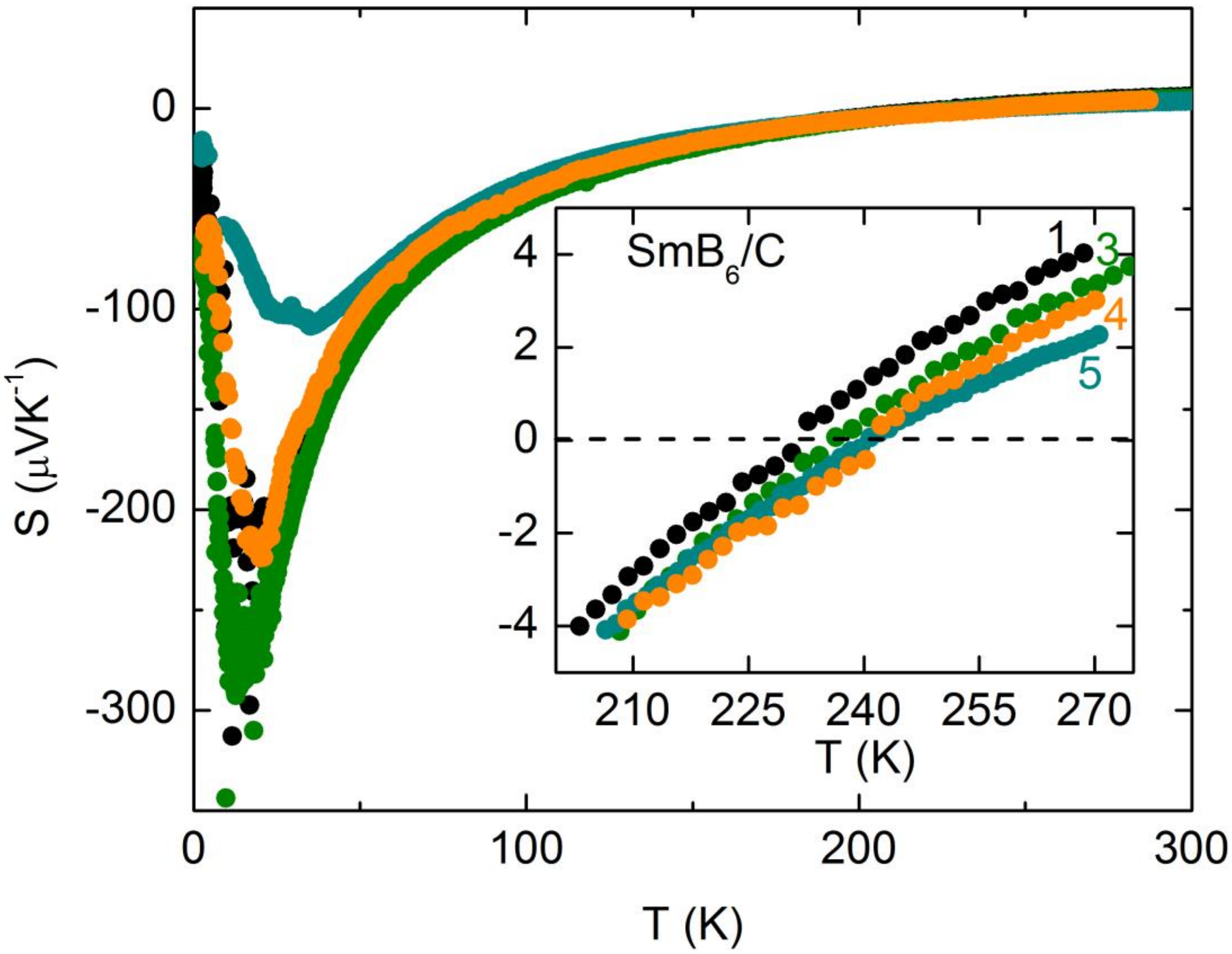

Fig. 4. (a) Magnetization data normalized by applied magnetic field (M/H) as a function of temperature for $SmB_6$ and $SmB_6$/C-1-5. (b) The Curie constant extracted from fits of the magnetization data from T = 100 –300 K for $SmB_6$ and $SmB_6$/C-1-5 as a function of $1/E_a$, a measure of the degree of plateauing of the low temperature resistivity, defined to be $1/slope_{2\text{-}5K}$ of $\ln(R/R_{300K})$ vs. 1/T. Tabulated values can be found in Table SII.

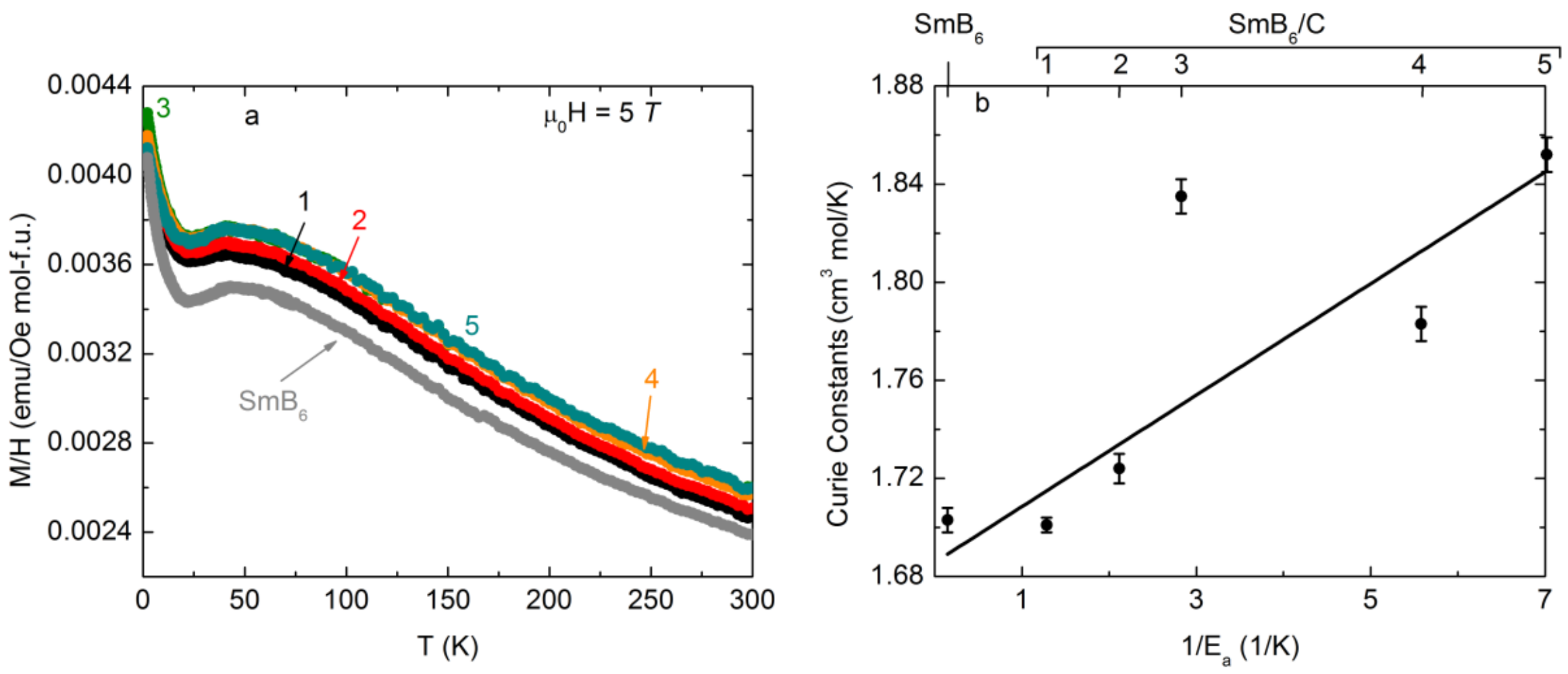

FIG. 5. (a) Specific heat divided by temperature ($C_p$ $T^{-1}$) vs. temperature for pure $SmB_6$, carbon-containing samples **2** and **5**, and the non-magnetic analog $LaB_6$. In addition to a low temperature (*ca*. T < 10 K) upturn that depends on carbon addition, there is a large excess specific heat that extends to room temperature. (b) Integration yields the changes in excess entropy; for all samples, the total is in excess of 2.3-2.9R (R = 8.314 J $K^{-1}$ mol-f.u.$^{-1}$).

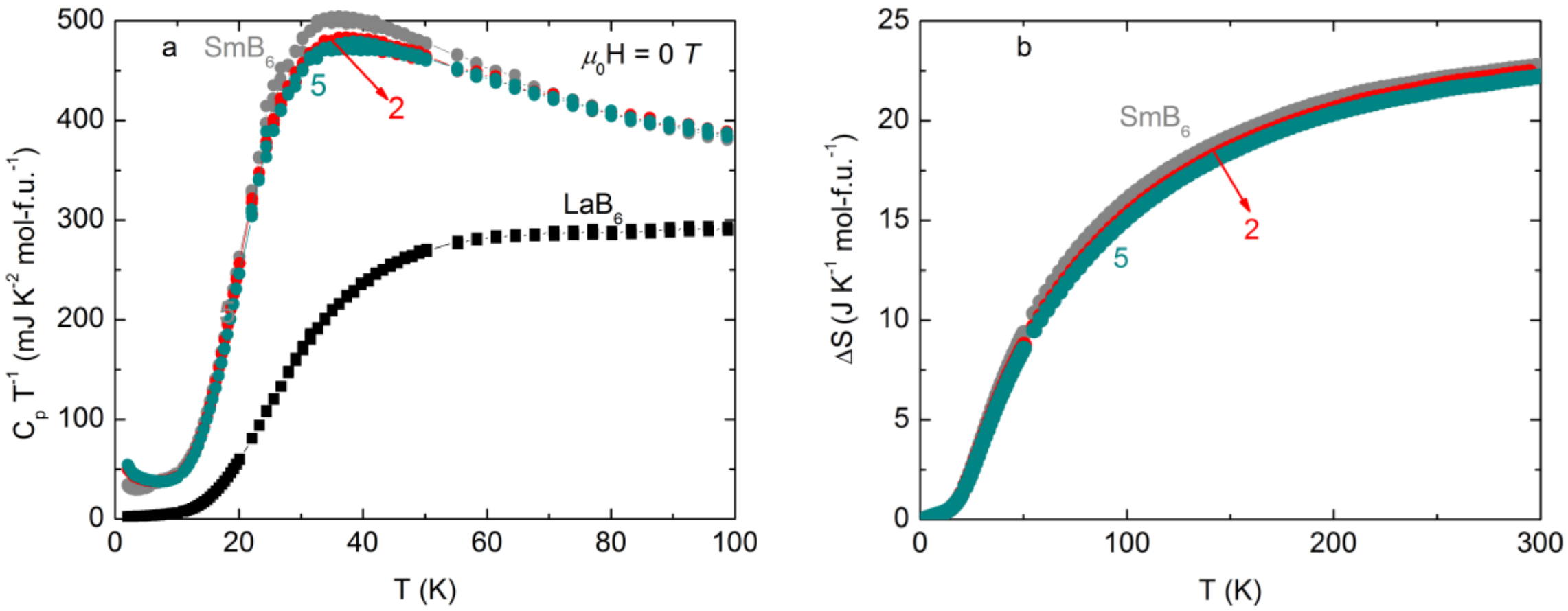

FIG. 6. (a) Specific heat divided by temperature for pure $SmB_6$ and the carbon-containing sample **5** under applied fields of $\mu_0H$ = 0 *T* (black), 3 *T* (red), 5 *T* (green), and 9 *T* (blue). Model fits to the data at T ≤ 10 K (red line) were performed using the expression $C_p T^{-1} = \gamma + \beta T^2 + AT^2\ln(T) + BT^{-3}$. (b) A plot of the β vs A coefficients shows that there is a universal linear relationship between the two parameters for all samples under all applied fields. The y-intercept corresponds to the lattice contribution to the specific heat, while the slope is given by –lnT* where T* = 17 K is a characteristic temperature for all samples and fields (see text).

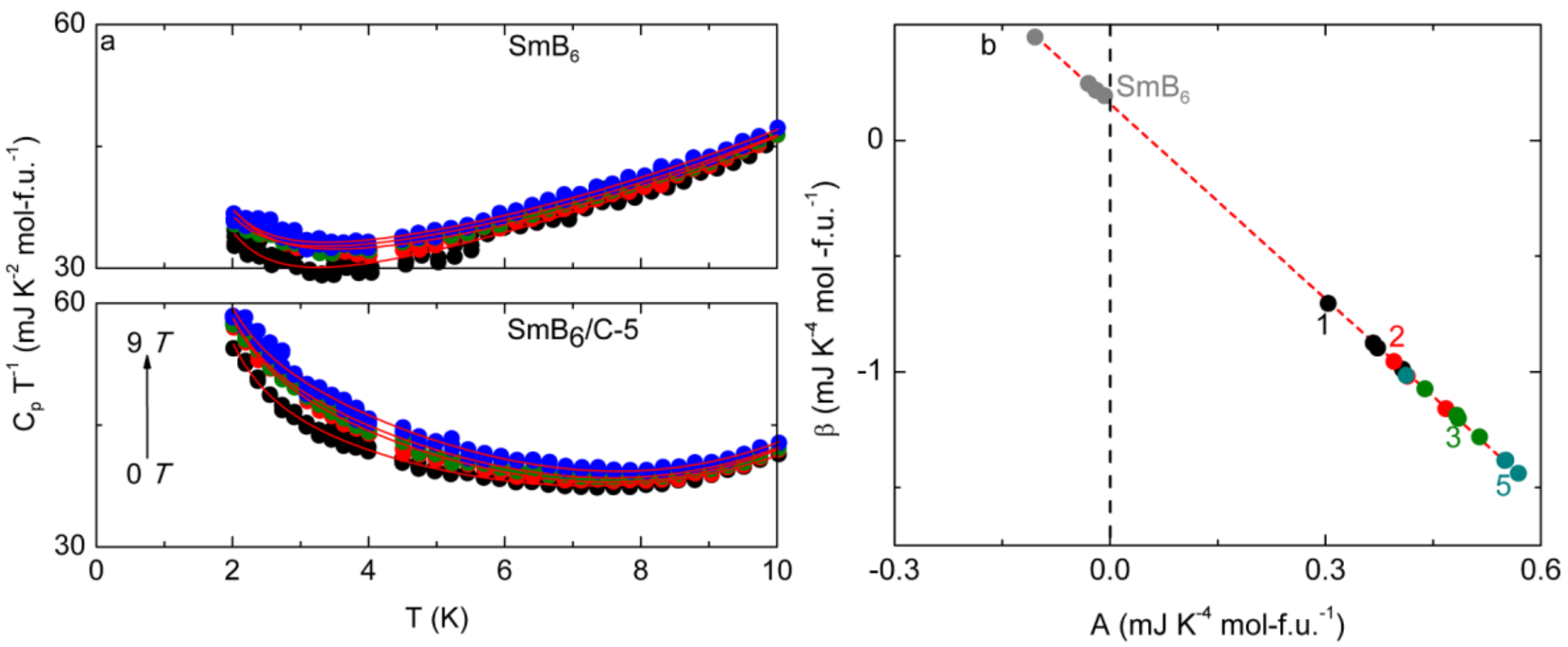

FIG. 7. (a) The negative to positive Seebeck crossover, $T_{s=0}$ (left), the Sommerfield coefficient $\gamma$, and temperature of the minimum value of the $C_p/T$ (right) increase concomitantly with $1/E_a$, a measure of the degree of plateauing of the low temperature resistivity. Error bars represent the statistical portion of the error (see Table SIII). (b) Schematic representation of the density of states of a Kondo insulator when $T \ll T_K$. The vertical lines show band filling positions estimated from the specific heat (see text). Our finding of an increase in low temperature conductivity (increased resistivity plateau) with naturally follows if there are surface states near the hybridization gap.

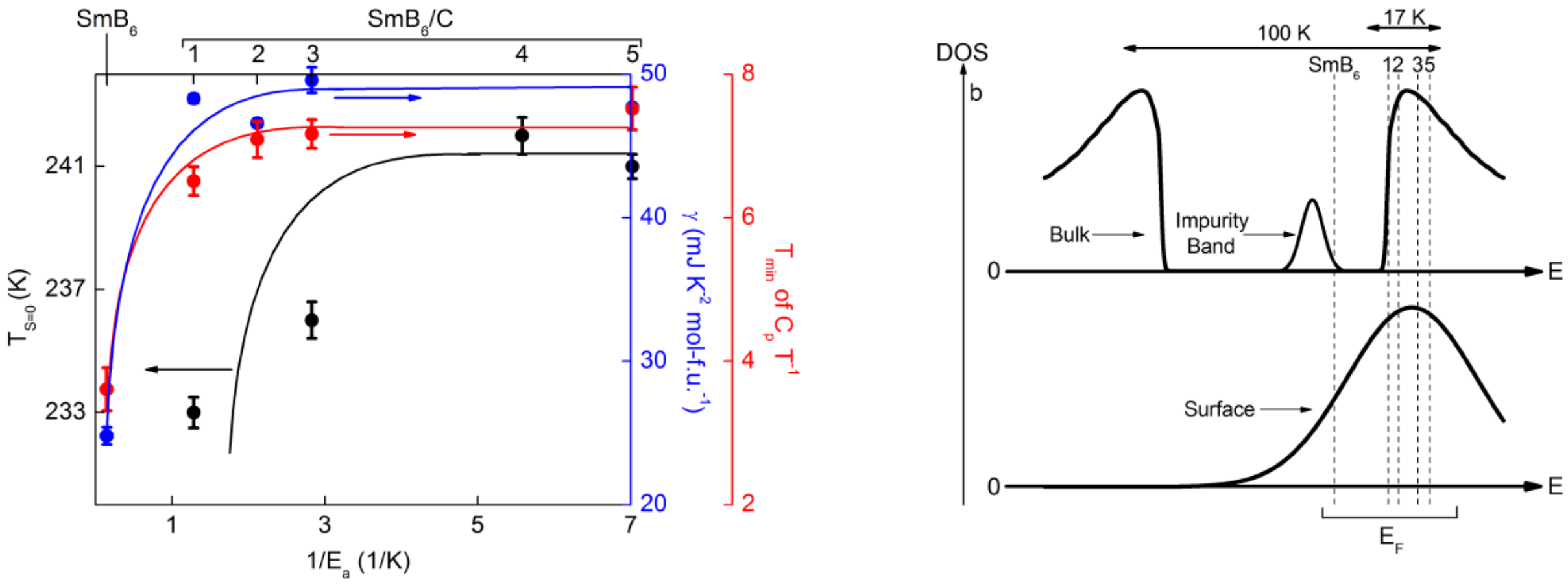

FIG. 8. (a) XAS and (b) XMCD spectra of surface and bulk $SmB_6$ collected at the Sm *M*-edge at T = 10 K and $\mu_0 H = 5$ *T*. The bulk is consistent with previous reports of a mixed valence mixture of $Sm^{2+}$ and $Sm^{3+}$ oxidation states (dashed lines). In contrast, the surface of $SmB_6$ contains entirely $Sm^{3+}$ (dashed lines). The surface displays a non-zero XMCD signal consistent with the presence of paramagnetic $Sm^{3+}$ ions.

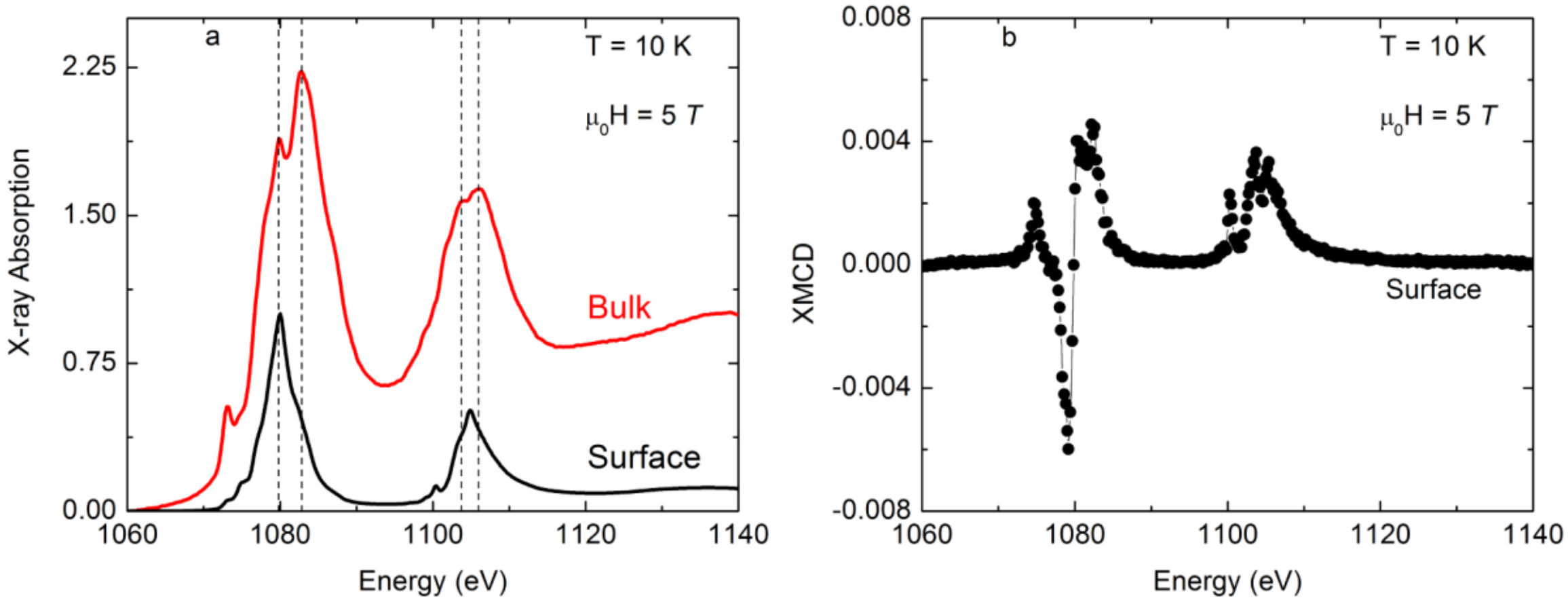

Correlation between bulk thermodynamic measurements and the low temperature resistance plateau in $SmB_6$

W. A. Phelan,[1,2] S. M. Koohpayeh,[2] P. Cottingham,[1,2] J. W. Freeland,[3] J. C. Leiner,[4]
C. L. Broholm,[2] and T. M. McQueen[1,2*]

[1]*Department of Chemistry, Johns Hopkins University, Baltimore, Maryland 21218, USA*
[2]*Institute for Quantum Matter, Department of Physics and Astronomy, Johns Hopkins University, Baltimore, Maryland 21218, USA*
[3]*Advanced Photon Source, Argonne National Laboratory, Argonne, Illinois 60439, USA*
[4]*Quantum Condensed Matter Division, Oak Ridge National Laboratory, Oak Ridge, Tennessee 37831, USA*

[*]*mcqueen@jhu.edu*

FIG. S1. Seebeck (S) coefficient vs. temperature for the carbon-containing sample $SmB_6$/C-4. The orange circles correspond to the original data set in the manuscript (FIG. 3.). The dataset represented by the black circles was collected after removing all original contacts of $SmB_6$/C-4, repolishing of the sample, and attaching of new electrical and thermal contacts. This demonstrates the reproducibility of our measurements of $T_{S=0}$.

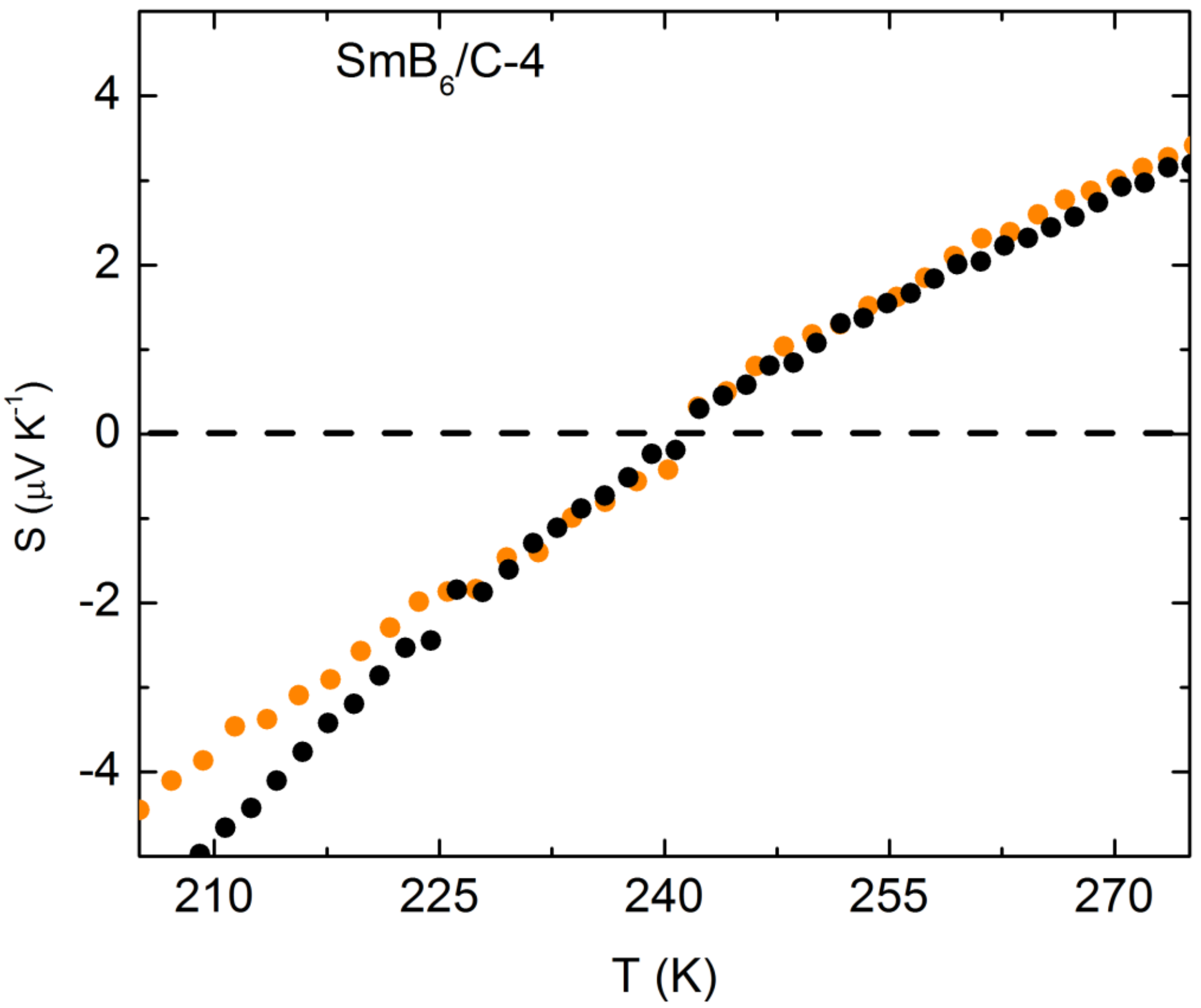

FIG. S2. Magnetization as a function of applied magnetic field at T = (a) 2 K, (b) 20 K, and (c) 200 K. The insets highlight the low applied magnetic field regions.

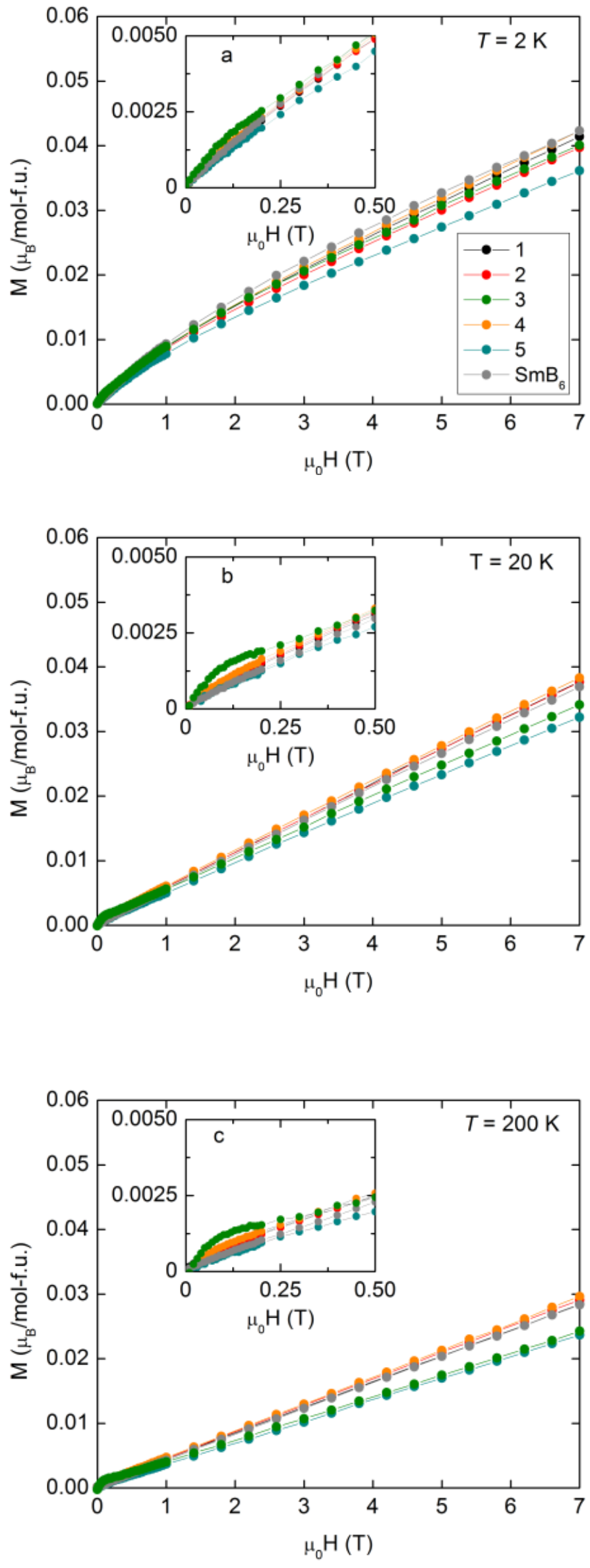

FIG. S3. Specific heat divided by temperature for $SmB_6$/C-5 at $\mu_0 H = 9$ *T*. The data $T \leq 10$ K were fit (shown in red) to the expression $C_p T^{-1} = \gamma + \beta T^2 + AT^2 \ln(T) + BT^{-3}$. The fit parameters can be found in Table 1 of the manuscript. The electronic, lattice, and Schottky contributions to the fit are shown as black, gray, and green curves, respectively.

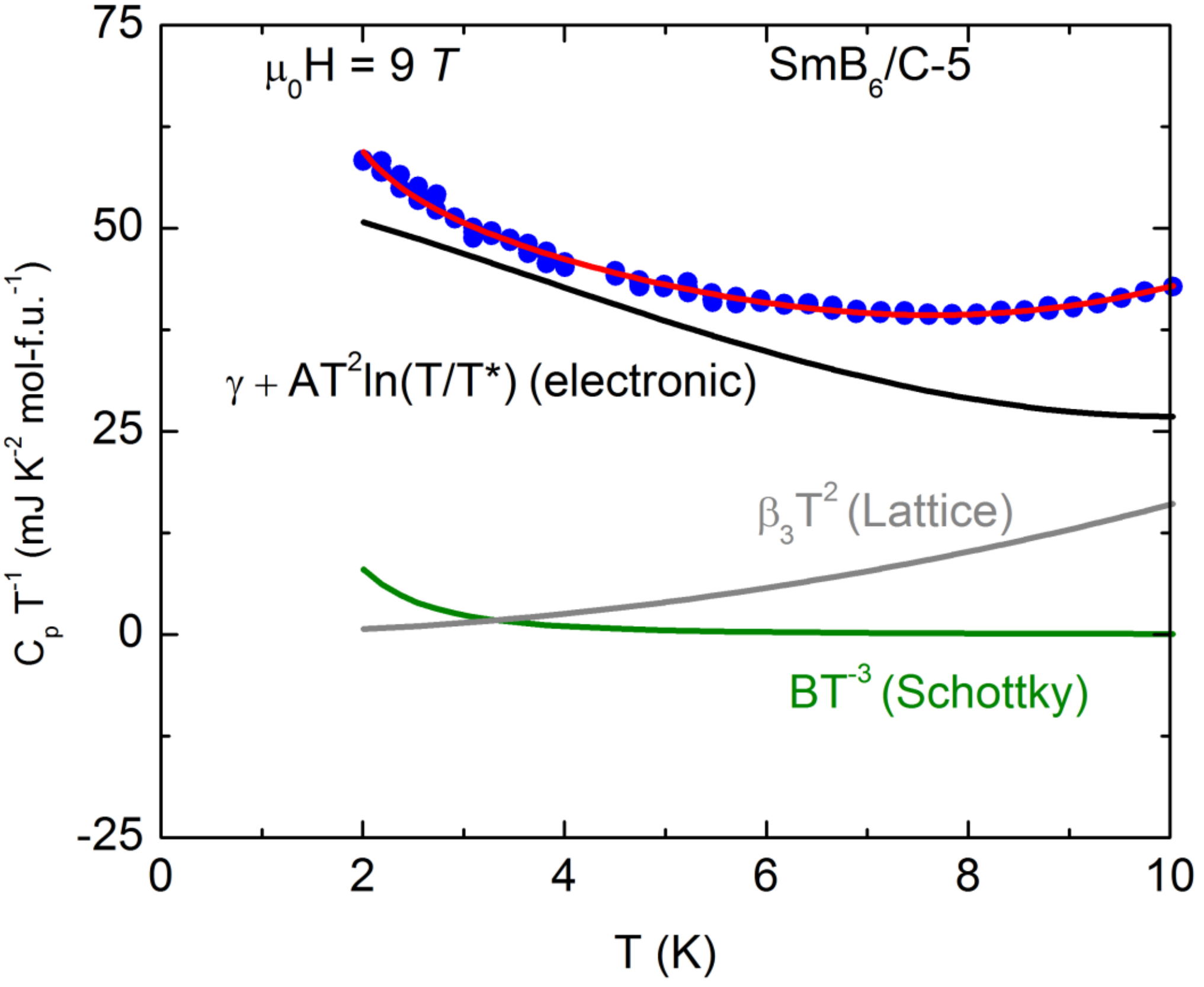

FIG. S4.  Activation energy plots of the normalized resistance (a) $SmB_6$/Al and (b) $SmB_6$/C.  The slope of the low temperature resistance changes systematically with carbon number, but not aluminum number.

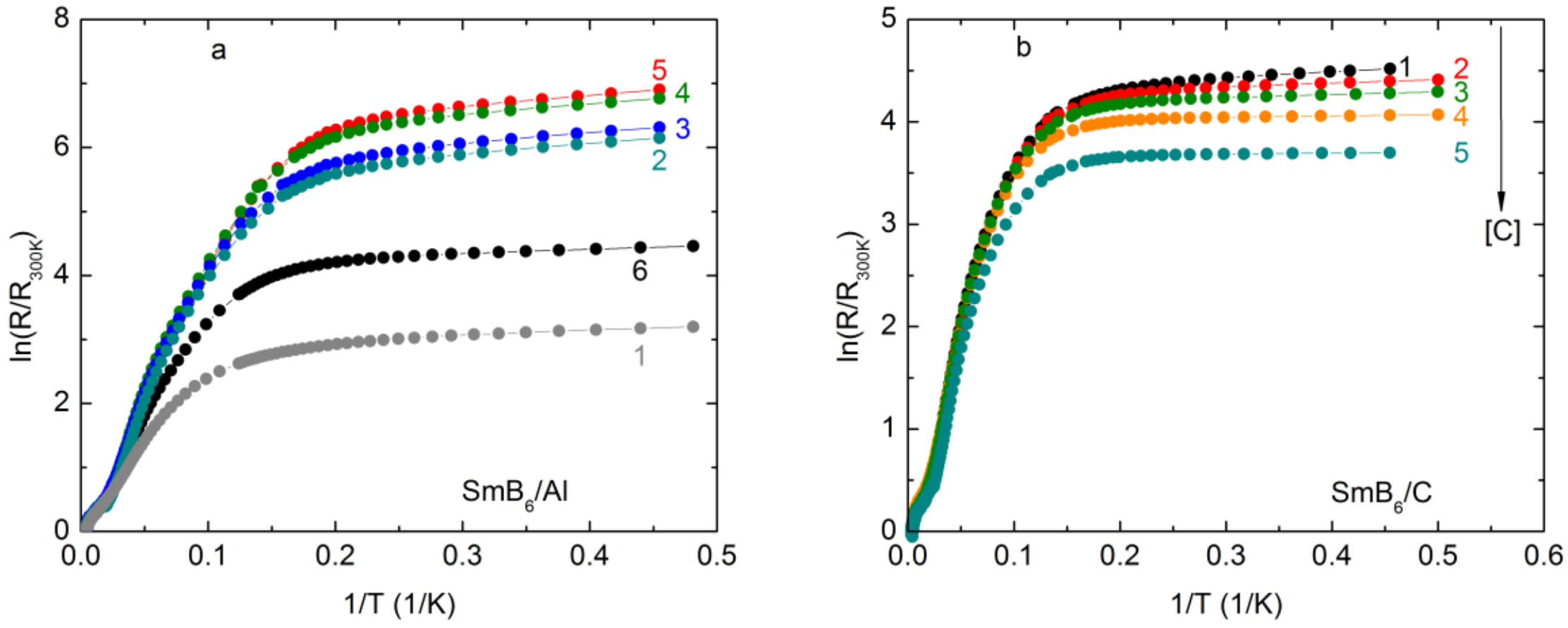

FIG. S5. (a) $R/R_{10K}$ versus temperature for $SmB_6$ sample measured at excitation currents of 50 µA (black) and 5000 µA (red). (b) Excitation current versus voltage for $SmB_6$ in the T = 1.95 – 10 K range. The I-V curves become progressively nonlinear with decreasing temperature. (c) The resistance divided by the resistance in the limit that the excitation goes to zero ($R/R_{I\to 0}$) versus applied power for two different $SmB_6$ samples at T = 1.95 K. The true resistance of the black curve is approximately five times greater that of the red curve, but the downturn occurs at similar applied power, implying a Joule heating effect rather than the "turn on" of a second conduction channel (e.g. sliding charge density wave).

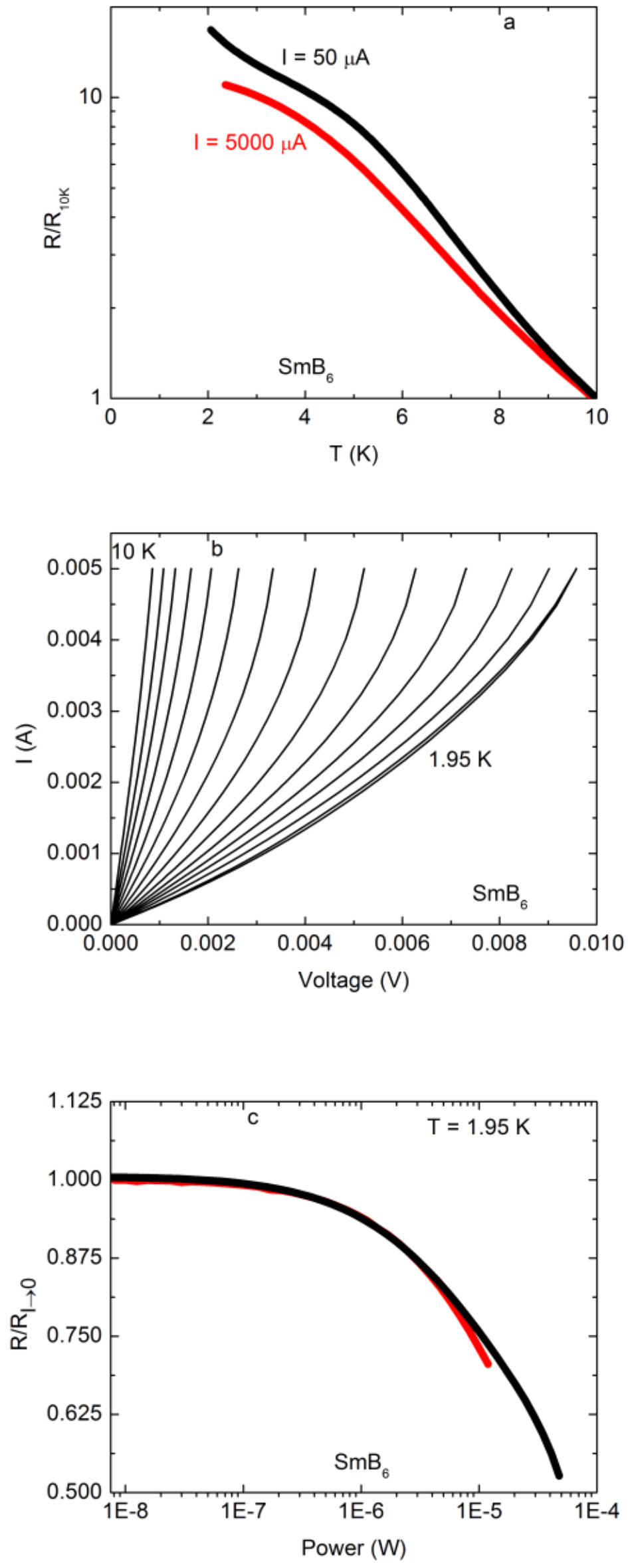

FIG. S6. As an additional check that the data presented in FIG. 2 are not contaminated by Joule heating effects, here we compare the measured resistance curves at the indicated excitation currents with the resistance values extracted from the IV curves extrapolated to the zero-current (no Joule heating) limit. The two methods of obtaining the resistances are in remarkable agreement, and imply Joule heating is not a significant contributor to the trends observed in the data in FIG. 2.

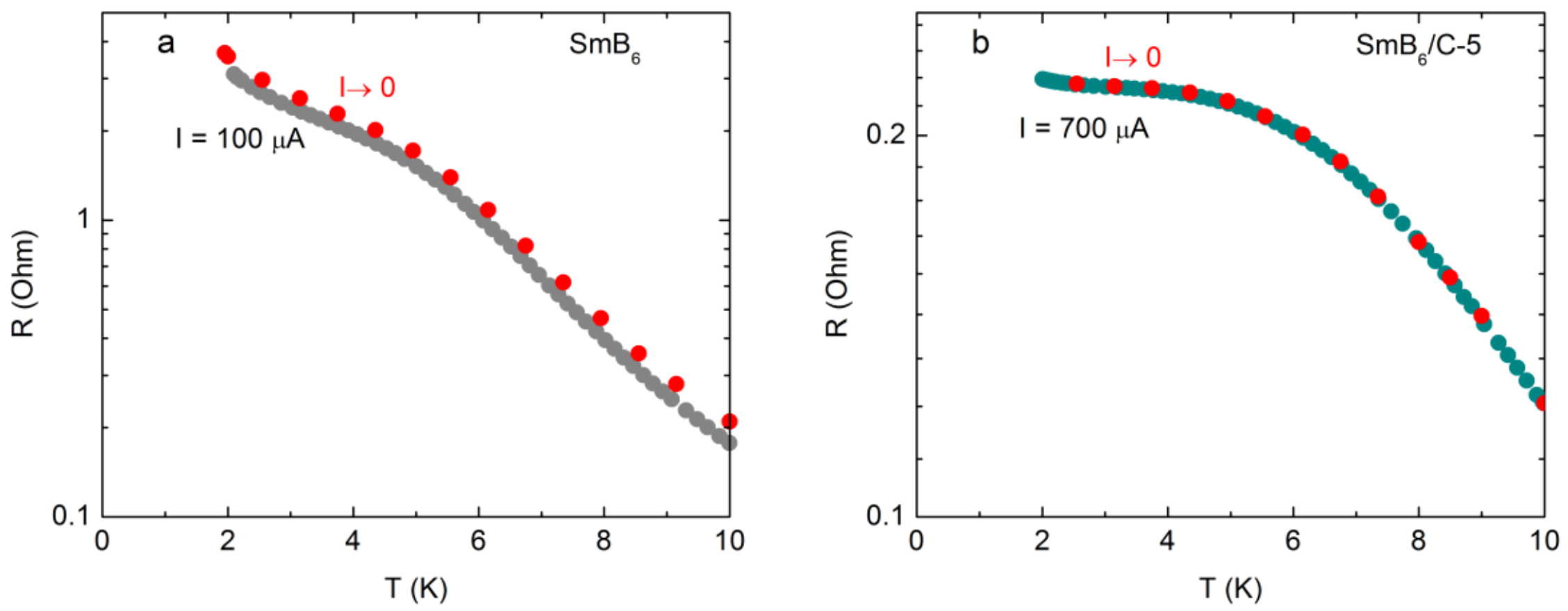

FIG. S7. Two-probe measurements to characterize the inner ($R_{23}$) and outer ($R_{14}$) contact resistances on representative specimens. These data imply contact resistances on the scale of R = 10-40 Ω, and suggest that both the sample and the contacts contribute to the Joule heating effects noted in Fig. S5.

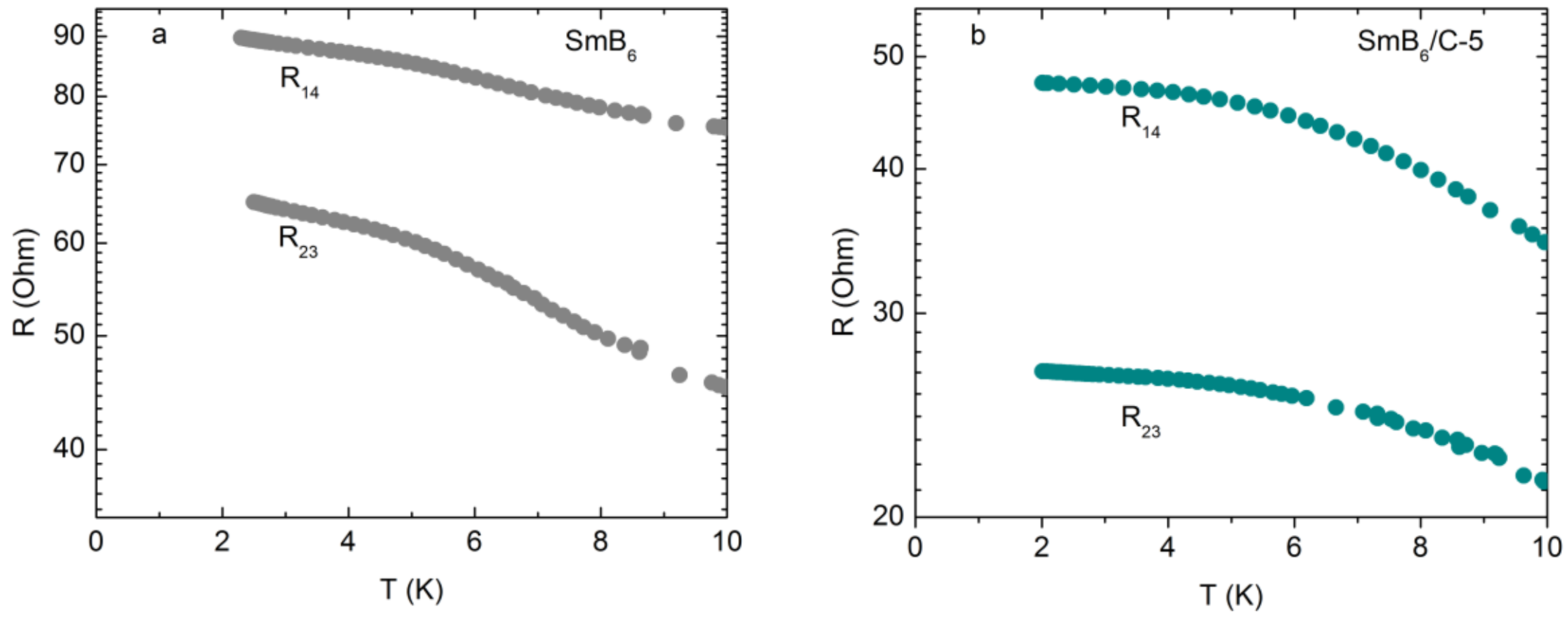

Table SI. Parameters extracted from fits to the resistance ($2 < T < 50$ K) of $SmB_6$, $SmB_6$/C-1-5, $SmB_6$/Al-1-6. Model fits to data were performed using the expression $R_1R_2/R_1+R_2 = (R_SR_Be^{\Delta 1/2T}e^{\Delta 2/2T})/(R_Se^{\Delta 1/2T} + R_Be^{\Delta 2/2T})$.

| | $R_S$ (Ω) | $R_B$ (Ω) | $\Delta_1$ (K) | $\Delta_2$ (K) |
|---|---|---|---|---|
| $SmB_6$ | 8(3) | 0.002(5) | 16(4) | 86(24) |
| $SmB_6$/C-1 | 0.0280(3) | 0.00048(2) | 1.68(6) | 82(1) |
| $SmB_6$/C-2 | 0.0742(7) | 0.00117(6) | 1.04(6) | 83(1) |
| $SmB_6$/C-3 | 0.0636(6) | 0.0011(5) | 0.80(6) | 85(1) |
| $SmB_6$/C-4 | 0.0488(4) | 0.00089(4) | 0.48(6) | 87(1) |
| $SmB_6$/C-5 | 0.063(7) | 0.00131(6) | 0.44(6) | 87(1) |
| $SmB_6$/Al-1 | 0.0748(7) | 0.0049(1) | 1.96(6) | 67(1) |
| $SmB_6$/Al-2 | 0.116(1) | 0.00084(5) | 3.60(6) | 74.6(8) |
| $SmB_6$/Al-3 | 0.0691(7) | 0.00048(3) | 3.56(6) | 75(1) |
| $SmB_6$/Al-4 | 0.0166(1) | 0.00081(5) | 3.50(4) | 74.4(8) |
| $SmB_6$/Al-5 | 0.268(2) | 0.00102(5) | 3.52(4) | 75.4(6) |
| $SmB_6$/Al-6 | 0.0367(3) | 0.00066(2) | 1.56(4) | 73.2(6) |

Table SII. Curie Constant extracted from linear fits to the slope of the H/M versus temperature data (100 < T < 300 K) of $SmB_6$, $SmB_6$/C-1-5.

| | Curie Constant ($cm^3$/K mol-f.u.) |
|---|---|
| $SmB_6$ | 1.703(5) |
| $SmB_6$/C-1 | 1.761(3) |
| $SmB_6$/C-2 | 1.724(6) |
| $SmB_6$/C-3 | 1.835(7) |
| $SmB_6$/C-4 | 1.783(7) |
| $SmB_6$/C-5 | 1.852(7) |

Table SIII. Parameters extracted from fits to the low temperature specific heat ($2 < T < 10$ K) of $SmB_6$, $SmB_6$/C-1, $SmB_6$/C-2, $SmB_6$/C-3, and $SmB_6$/C-5 under various applied magnetic fields. Model fits to the specific heat data were performed using the expression $C_p/T = \gamma + \beta T^2 + AT^2 \ln(T) + BT^{-3}$. The individual terms are defined in the text. An example fit is shown in Fig. S3. Test fits of the pure and SmB6/C-5 samples over different temperature ranges (T = 3 to 12 K, 2 to 12 K) show that while there is some variability in the precise values of ß and A depending on the range used, they still follow the universal relationship presented Fig. 6, and the values of γ change by at most 10-15%.

| | Field | γ (mJ $K^2$ mol-f.u$^{-1}$) | β (mJ $K^{-4}$ mol-f.u$^{-1}$) | A (mJ $K^{-4}$ mol-f.u$^{-1}$) | B (mJ K mol-f.u $^{-1}$) |
|---|---|---|---|---|---|
| $SmB_6$ | 0 *T* | 24.8(6) | 0.45(6) | -0.10(3) | 66(5) |
| | 3 *T* | 28.6(3) | 0.19(3) | -0.01(1) | 65(3) |
| | 5 *T* | 29.2(3) | 0.22(3) | -0.02(1) | 52(3) |
| | 9 *T* | 29.4(4) | 0.25(4) | -0.03(2) | 56(3) |
| $SmB_6$/C-1 | 0 *T* | 48.3(3) | -0.70(3) | 0.30(1) | 40(2) |
| | 3 *T* | 51.3(2) | -0.88(2) | 0.37(1) | 44(2) |
| | 5 *T* | 52.4(2) | -0.90(2) | 0.37(1) | 35(2) |
| | 9 *T* | 54.4(2) | -0.99(2) | 0.41(1) | 33(2) |
| $SmB_6$/C-2 | 0 *T* | 46.6(3) | -0.96(3) | 0.40(1) | 55(3) |
| | 3 *T* | 49.0(4) | -1.07(4) | 0.44(2) | 64(4) |
| | 5 *T* | 49.2(4) | -1.02(4) | 0.41(2) | 65(4) |
| | 9 *T* | 51.7(4) | -1.16(5) | 0.47(2) | 59(4) |
| $SmB_6$/C-3 | 0 *T* | 49.6(9) | -1.07(9) | 0.44(3) | 59(7) |
| | 3 *T* | 51.7(3) | -1.19(3) | 0.48(1) | 67(2) |
| | 5 *T* | 52.7(3) | -1.20(3) | 0.49(1) | 58(2) |
| | 9 *T* | 54.7(3) | -1.28(3) | 0.51(1) | 56(2) |
| $SmB_6$/C-5 | 0 *T* | 47.7(2) | -1.02(3) | 0.41(1) | 83(2) |
| | 3 *T* | 53.0(3) | -1.38(4) | 0.55(1) | 74(3) |
| | 5 *T* | 53.8(3) | -1.38(3) | 0.55(1) | 68(2) |
| | 9 *T* | 55.6(4) | -1.44(5) | 0.57(2) | 64(4) |

Extraction of Magnetic Moment Information from Surface XMCD

The extraction of magnetic moment information from the surface XMCD data was performed following the same procedure as that reported by Dhesi, et al. for $SmAl_2$ [1]. A linear background was removed from the XAS spectrum prior to integration. We assumed that the $X_I/X_E$ ratio [2] for the MCD sum rule was 3.0, and that the ratio $\langle T_z \rangle / \langle S_z \rangle = -0.2$ [3].

Supplementary References